\documentclass[letterpaper,twocolumn,10pt]{article}
\usepackage{usenix2025_SOUPS}

\usepackage{tikz}
\usepackage{amsmath}

\usepackage{filecontents}

\begin{document}

\date{}

\title{\Large \bf Examining User Behavior and Cognitive Biases in Personal Password Security}

\def\plainauthor{Author name(s) for PDF metadata. Don't forget to anonymize for submission!}

\author{
{\rm Evelyn Crowe}\\
\small Texas A\&M University
\and
{\rm Patralika Ghosh}\\
 \small Texas A\&M University
\and
{\rm Shreyas Kumar}\\
\small Texas A\&M University
\and
{\rm Rebecca Schlegel}\\
\small Texas A\&M University
\and
{\rm Rebecca Ward}\\
\small Texas A\&M University
\and
{\rm Guofei Gu}\\
\small Texas A\&M University
} 

\maketitle

\begin{abstract}
Despite increasing awareness of cybersecurity risks, users continue to engage in insecure password practices, such as reusing passwords, choosing weak credentials, and neglecting security recommendations. The study explores the behavioral and cognitive factors that influence password decision-making by integrating insights from behavioral economics, particularly hyperbolic discounting, status quo bias, and present bias. We conducted a survey to analyze how people create, store and manage their passwords, examining whether security habits have improved over time in response to greater awareness. Our findings reveal that immediate convenience often outweighs long-term security considerations, leading users to prioritize memorability over strength. Additionally, we identify key psychological biases that contribute to security procrastination and resistance to adopting more secure authentication practices, such as password managers and multi-factor authentication. The study contributes to human-centered security by bridging the gap between security awareness and action, offering practical insights to design user-friendly authentication policies that align with real-world decision-making tendencies. \\
\end{abstract}
\section{Introduction}
As the internet continues to expand, cybercrime is rising in parallel. According to the Federal Bureau of Investigation's (FBI) Internet Crime Report, the Internet Crime Complaint Center (IC3) received 1,000,597 complaints in 2025, compared to 859,532 complaints in 2024, along with an estimated 26\% increase in reported losses in 2025 \cite{fbi_ic3_2024, fbi_ic3_2025}. IC3 reported over 3,600 ransomware complaints, with losses exceeding \$32 million in 2025. To mitigate such threats, FBI suggestions include eliminating default credentials, enforcing NIST password standards, and implementing multi-factor authentication (MFA), particularly for high-risk systems, in conjunction with other non-password-based security controls \cite{fbi_ic3_2025}. These recommendations underscore the critical role of authentication security in mitigating cyber threats. \\

Despite usability and security concerns, passwords remain a dominant form of authentication. Although alternative authentication methods, such as token devices (e.g, NFC cards, USB tokens), biometric authentication, and two-factor authentication, provide varying degrees of security \cite{guven2022novel}, text-based passwords remain the dominant form of authentication \cite{barkadehi2018authentication, herley2011research}. However, prior research indicates that users frequently rely on predictable password creation strategies, often combining multiple approaches, such as using names, dictionary words, and appending numbers or symbols to the beginning or end of a word\cite{219404}. In 2025, a study analyzing the lifecycle of leaked authentication credentials reported a significant password reuse rate of 72.5\%. The study further found that commonly used passwords exhibit predictable patterns, including repeated sequences, leetspeak substitutions, keywalking patterns, and alphanumeric combinations. \cite{rabzelj2025beyond}. \\

Experts frequently advocate for password managers to improve security, integrating secure password storage, retrieval, and randomized password generation. These tools enhance security and usability by mitigating the risks of weak or reused passwords. However, adoption remains low, as many users struggle to understand how password managers function and express concerns regarding their security and trustworthiness \cite{tavakol2011making, pearman2019people}. Research has found that individuals who did not use password managers relied on methods such as memorization, written notes, self-sent emails or voicemails, unencrypted files, and note-taking applications to manage passwords, with many also depending on password reset mechanisms for recovery \cite{pearman2019people, aurigemma2017so}.\\

A key issue in password security is the gap between users’ awareness of recommended practices and their actual behavior. A similar pattern has been observed in the context of privacy: although users often report that they value privacy, they can still disclose personal information or take limited action to protect it \cite{kokolakis2017privacy, barth2017privacy}. Users express security concerns, but engage in behaviors that contradict these concerns, such as reusing passwords across multiple accounts or neglecting to update old credentials. To better understand these behaviors, the study examines behavioral economics concepts such as hyperbolic discounting, status quo bias, and present bias, which help explain why users often make poor security decisions. Hyperbolic discounting explains why users prioritize short-term convenience over long-term security. Users are more likely to choose easily memorable passwords or reuse existing ones because the effort required to create and manage strong passwords is immediate, while the risk of a breach feels distant \cite{laibson1997golden}. Status quo bias suggests that users tend to stick with default or familiar security habits rather than adopting more secure alternatives, even when there are better options. Resistance to change can explain why users hesitate to adopt password managers or enable multi-factor authentication \cite{samuelson1988status, baumeister2001bad}. Present bias contributes to security procrastination, where users defer security-enhancing behaviors, such as updating passwords or enabling multi-factor authentication, in favor of immediate convenience. Research has shown that commitment devices and security nudges can help mitigate these tendencies by encouraging proactive security actions \cite{frik2018better, o1999doing}. Our study explores the intersection of password security, user behavior, and behavioral economics to gain deeper insights into the decision-making processes that shape personal password practices.\\

Our study examines what factors shape the ways individuals choose and manage their passwords. To address these questions, we examine the disconnect between users’ knowledge of password security best practices and their actual password behaviors. To support our analysis, we conducted a large-scale survey with 248 participants, examining their password habits and security awareness. The study analyzes the gap between security knowledge and real-world behavior, shedding light on why users fail to follow best practices despite understanding their importance. \\

\section{Background}
In this section, we introduce key behavioral economic models, including hyperbolic discounting, status quo bias, and present bias, to explain the cognitive biases underlying decision-making in password security. \\

Hyperbolic discounting describes the tendency of individuals to devalue future rewards or consequences in favor of immediate gratification \cite{laibson1997golden}. People do not apply a uniform or a “constant” discount rate over time. Instead, they discount future rewards more steeply when those rewards are closer to the present and less steeply when they are farther away. In other words, the difference between getting something right now versus in one week feels larger and is discounted more heavily than the difference between getting something in 100 weeks versus 101 weeks. Consequently, the “implicit discount rate,” which is the rate at which someone devalues future benefits, is higher in the short term but lower in the long term\cite{frederick2002time}. When applied to password security, the phenomenon explains why users prioritize the ease of remembering a simple password over the distant risk of a data breach. A strong password provides long-term protection, but the mental effort required to create and recall the password is immediate. \\

Samuelson and Zeckhauser show that people often stick with the option they already have, or the default choice, even when another option may be better. The authors define status quo bias as a systematic preference for maintaining a current decision or situation. The bias leads individuals to do nothing or stick with their existing option instead of changing, even when switching could yield more significant benefits\cite{samuelson1988status}. Switching from the status quo involves potential losses and gains, but losses are typically felt more acutely than gains\cite{baumeister2001bad}. But, maintaining the status quo requires less mental effort than evaluating and justifying a new choice, and people may fear regretting a bad switch more than missing out on a good one. When outcomes are uncertain, sticking to what is familiar feels safer, even if it is not necessarily optimal \cite{samuelson1988status}. \\

People are often present-biased, meaning they favor immediate gratification over future rewards \cite{laibson1997golden}. The preference for time leads individuals to prioritize immediate benefits and defer activities that involve upfront costs \cite{o1999doing}. As a result, they tend to procrastinate tasks requiring effort while fast-tracking those offering immediate rewards. A relevant study explored the impact of the present bias on the user adoption of essential computer security tasks, such as enabling automatic updates, restarting devices, and enrolling in two-factor authentication. Present bias often leads users to neglect these security measures. The researchers investigated the application of commitment devices, a behavioral economic technique, to mitigate the effects of present bias. Through two online experiments involving more than 1,000 participants, they found that both reminders and commitment nudges effectively reduce user intentions to ignore security-related actions, particularly by enabling automatic updates and adopting multi-factor authentication. However, these nudges proved less effective for configuring automatic backups \cite{frik2018better}. 

\section{Related Works}
The section summarizes previous work on user password habits and management choices. We also examine studies that have explored the adoption of password managers, as well as problems with password managers that might hinder their effective use.

\subsection{Password Habits}
Text-based passwords are the most common form of authentication \cite{barkadehi2018authentication, herley2011research}. Bounded rationality, misconceptions about risk, and users' desire for memorability are the some of the factors contributing to poor password habits. Bounded rationality becomes apparent when comparing users' self-reported password habits with their actual behavior\cite{hanamsagar2018leveraging}. Users often underestimate the number of accounts they have, describe more rational reuse strategies than they practice, and report different password-composition strategies that they employ \cite{hanamsagar2018leveraging}. Most people create passwords using their own strategies, often developed independently. The less secure algorithms frequently involve reusing a base string with minor variations across different accounts or reusing entire passwords verbatim\cite{ur2015added,219404,pearman2017let}. Some sought advice from articles or security-training classes. Regardless of the method, many hold misconceptions about password security \cite{ur2015added,ur2016users}. People vastly overestimate the security benefit of adding digits to passwords while underestimating the predictability of keyboard patterns and common phrases \cite{ur2016users}. However, when people follow password creation policies, they tend to surpass the required length and character class requirements, resulting in average passwords that are significantly more complex than necessary, suggesting that many users voluntarily create more secure passwords rather than just meeting policy constraints \cite{wash2021prioritizing,ur2015added,shay2014can}. Hanamsagar et al. highlighted a significant lack of awareness among users regarding the risks and capabilities of attackers in their study. The study revealed that many users were poorly informed, with some unaware of password-reuse attacks and others mistakenly believing that strong passwords were immune to such attacks. Additionally, a notable portion of users deemed the likelihood of these attacks to be low \cite{hanamsagar2018leveraging}. A study by Shay et al. found that users in their research perceived the following type of password to be the most secure: "Passwords in this condition must include at least eight characters, including a lowercase English letter, an uppercase English letter, a digit, and a symbol (something that is not a digit or an English letter)" \cite{shay2014can}. \\

People often prioritize creating passwords that are easy to remember.\cite{219404,choong2014united,wash2016understanding,hanamsagar2018leveraging}. People predominantly rely on memorization for recalling their passwords, using the "Change Password" option as a backup \cite{219404}. However, many people also depend on storing their passwords on paper or electronically \cite{choong2014united,stobert2018password,shay2010encountering, komanduri2011passwords}. Relying on the "Change Password" option has its shortcomings, as many participants reported experiencing at least one account lockout in the past year, with some encountering three or more lockouts. \cite{choong2014united,219404}. People often reuse passwords, so when they forget a password for a particular website, they typically go through all their possible passwords until they find the right one \cite{ion2015no,wash2016understanding}. A study conducted by Habib et al. suggests that forced password expiration may not have the anticipated negative or positive effects. Participants generally employed one or two methods to update their passwords. Most often, they modified their previous password, typically by capitalizing a letter \cite{219404,zhang2010security}. But, fewer participants reused passwords from other accounts, while a notable portion created an entirely new password for their main workplace account or used a password generator \cite{219404,shay2010encountering}. Recent research has examined the effectiveness of various password policies to enhance password security while maintaining user convenience. The findings suggest that more effective strategies include mandating passwords to be at least eight characters long and prohibiting those easily guessed within a limited number of attempts. These properly configured minimum-strength policies enhance security, particularly against extensive offline attacks, and preserve usability. Although password policies effectively enhance password strength, people often fill them predictably \cite{tan2020practical}. However, implementing complex dictionary word checks in password creation policies has been associated with decreased usability, making the password creation process more difficult \cite{komanduri2011passwords}.\\

\subsection{Password Management}
Research investigating people's current password habits and challenges offers valuable insights into their password-management decisions. Password managers make password management easier, but many people avoid using them because they do not fully understand how they work or the security measures behind them \cite{pearman2019people,stobert2018password}. People create passwords using their custom algorithms, often developed independently. Most do not have a dominant password used across most websites but have multiple passwords that they reuse \cite{wash2021prioritizing,219404,wash2016understanding}. Research suggests that password reuse remains common even among some password manager users \cite{pearman2017let,wash2016understanding}. A recent study by Pearman et al. discovered that individuals utilizing independently installed password managers typically manage anywhere from 50 to nearly 1000 password-protected accounts\cite{pearman2019people}. \\

The study revealed that individuals who abstain from using password management tools tend to exhibit risky password behaviors, such as password reuse and a scarcity of unique passwords\cite{pearman2019people}. Participants who utilized independently installed password management tools gradually shifted away from their old habit of password reuse, opting for auto-generated unique passwords after they experienced phishing attacks targeting their reused passwords \cite{pearman2019people}. A study examined how people alter their passwords following a data breach. Approximately one-third of participants with compromised accounts changed their passwords on the breached domain, with only a tiny fraction doing so within three months. On average, participants had many similar passwords but changed only a few within a month after updating the breached password. Most changes resulted in weaker or equally strong passwords. Regardless of changes to similar passwords within a month, new passwords on the breached domains remained similar to the participants' other passwords \cite{bhagavatula2020people}. Despite the growing need for password management tools, many people still avoid using them. Studies found that individuals often prefer memorizing reused passwords or storing them in handwritten notes or their phone's notes app for easy access \cite{choong2014united,219404,pearman2019people,stobert2018password}. On the other hand, in some studies people have faced difficulties recalling slight variations of reused passwords and were concerned about the security of physical or digital storage methods. Many participants said they did not use password managers because they were unfamiliar with them or unsure about how secure and reliable they were. Some felt they did not have important or sensitive enough information to need a password manager. Others worried that storing all passwords in one place could create a single point of failure if the password manager was ever compromised. In addition, a few participants had negative past experiences with password managers, which made them trust these tools less.\cite{pearman2019people,stobert2018password,gao2018forgetting,luevanos2017analysis}.\\

\section{Methodology}

Our study investigates the disconnect between password security awareness and actual password practices using  survey responses with 248 college students. We aimed to explore:
\begin{description}
    \item[RQ1:] How does awareness of password security influence personal password habits ?
    \item[RQ2:] What factors influence how individuals choose and manage their passwords?
\end{description}

The study engaged a diverse group of 248 collegiate individuals aged between 18 and 25. The survey contained an array of 35 questions, including both open-ended and multiple-choice formats. The demographic breakdown of the participants included 153 females (61.7 \%), 93 males (37.5 \%), 1 individual identifying as non-binary or third gender(0.4 \%), and 1 person preferred not to reveal their gender(0.4 \%). The participant pool showcased a significant representation of the Science, Engineering, Medicine, Law, Business, and Social Science disciplines. The survey was structured into three sections for a thorough understanding of the participants' backgrounds and behaviors. The initial segment gathered foundational data, focusing on demographic details and educational background. The second segment gathered information of their personal password creation and management habits. The concluding segment focused on the participant's awareness and knowledge levels regarding optimal security protocols and password practices. \\

The survey consisted of two types of questions: multiple-choice questions and open-ended questions, allowing participants to elaborate on their reasoning and thought process behind specific decisions. To analyze and categorize open-ended responses, we employed a thematic coding approach to identify patterns and recurring themes in participant explanations. Responses were first grouped based on their underlying rationale, such as memorability, security concerns, personal significance, or external requirements such as system-enforced constraints. The analysis allowed us to quantify key trends, such as the prevalence of security-conscious behavior versus convenience-driven choices. The insights derived from the open-ended questions helped contextualize user behavior beyond multiple-choice responses, revealing nuanced decision-making processes that influence real-world password practices. The survey was conducted with the approval of our university's Institutional Review Board (IRB). No personal information was collected to protect participant privacy, and all responses were recorded anonymously. Data protection guidelines, as advised by the IRB, were followed. \\

\section{Results}
The section presents the findings from our survey, which examined participants' password creation strategies, their awareness of security best practices, and their actual behaviors in managing passwords. \\

\subsection{Setting Passwords and Password Length}
According to NIST Special Publication 800-63B, the recommendation given to password verifiers is that passwords used as a single-factor authentication system should have a minimum length of 15 characters. However, when passwords are combined with additional authentication factors in a multi-factor authentication system, the minimum recommended length can be reduced to eight characters \cite{nist80063b}. We surveyed participants about their typical password lengths to examine real-world password habits. The results showed that 127 participants (51.2\%) use passwords of length 12, 78 participants (31.4\%) use passwords of length 8, and 20 participants (8.0\%) indicated that their password length varies depending on its purpose.
Additionally, 19 participants (7.6\%) reported using 17-character passwords, while two participants (0.8\%) used passwords of 22 characters. Another two participants (0.8\%) selected "Other," specifying 6 and 10 characters lengths. To assess knowledge of best security practices, we also asked participants how long they believe a password should be. The majority (144 participants, 58.1\%) recommended a length between 8-12 characters, while 88 participants (35.1\%) favored a range of 13-20 characters. Only one respondent (0.4\%) suggested 21-32 characters, while 8 participants (3.2\%) advocated for 33+ characters, and another 8 participants (3.2\%) stated that length does not matter.\\

\subsection{Common Patterns in Passwords}
A study by Shay et al. found that nearly 80\% of their users based passwords on a word or name, often adding special characters at the beginning or end \cite{shay2010encountering}. In our study, 179 participants (72.1\%) recognize that using personal information in passwords is unsafe and 19 participants (7.6\%) believe that the use of personal information in passwords can be safe. However, 142 participants (57.2\%) admitted including personal information in their passwords, 31 participants (12.5\%) acknowledged that they occasionally use personal information in their passwords and 75 participants (30.2\%) said that they never use personal information in their passwords. Our survey asked why participants use personal information and most reported prioritizing memorability and convenience over security. 50 participants (20.1\%) reported that they believe using personal information in passwords can be safe in certain situations, depending on different factors. When asked to elaborate, they explained that they are more likely to use highly personal information, unlikely to be found online, easy to remember, or related to the type and importance of the account. We asked our participants if they use words from the English dictionary and 100 participants (40.3\%) said yes, 35 participants (14.1\%) said sometimes and 113 participants (45.5\%) said no.\\

Among the participants who responded “yes”, the most common reason was the ease of memorability (65.0\%). Other reasons included the belief that dictionary words are required for password validity (7.0\%), personal significance of the words (16.0\%), passwords being set by parents and never changed (2.0\%), a lack of specific reasoning (4.0\%), the belief that using a dictionary word makes the password more unique (3.0\%), and a preference for using phrases as passwords (2.0\%). In contrast, participants who responded “no”, the majority (77.2\%) reported keeping personal passwords that are easy to recall. Additionally, 13 participants said they use names of important people, 6.1\% exclusively use non-dictionary numbers or letters, and 16.7\% avoid common words for added security. Among the participants who said "sometimes", 14.3\% said it depends on personal significance, 22.9\% stated it is necessary in some cases, 25.7\% said they choose words randomly, 28.6\% said their choice depends on the website they are using, and 8.6\% said they prefer simpler passwords for accounts they use daily. Overall, the findings suggest that while participants understand password security concepts, their choices are heavily influenced by convenience and context. High-security accounts tend to have more complex passwords, whereas lower-security accounts often have simpler passwords. We asked whether the participants think its good practice to use words from the English dictionary in passwords, 48 participants (19.3\%) responded “yes”, 87 participants (35.4\%) responded “no” and 110 participants (44.3\%) stated that "it depends". A similar result was observed in Hanamsagar et al.'s study, where 93\% of participants used names and words of personal significance in their passwords. However, to increase strength, they added numbers, symbols, and capitalized parts of their passwords \cite{hanamsagar2018leveraging}. Our participants argue that specific, obscure personal details can enhance memorability without significantly increasing risk. The participants in our study appear to take a conditional approach by balancing security and convenience based on the type of account and the kind of personal information used in the password.  \\
 
\subsection{Changing and Reusing Passwords}
A study by NIST found that more than 70\% of participants preferred that passwords remain valid for more than 90 days before requiring a change \cite{choong2014united}. Our study produced similar results, revealing that a significant majority (157 participants, 63.3\%) only changed their passwords when required, 59 participants (23.8\%) admitted to never changing their passwords, 11 participants (4.4\%) said that they change their passwords every six months, 8 participants (3.2\%) said that they change their passwords every three months, 9 participants (3.62\%) said that they change their passwords once every year and 4 participants (1.6\%) said that they change their passwords every month. When participants were asked how often passwords should ideally be changed, many believed that passwords should be updated annually (63 participants, 25.4\%), every six months (57 participants, 23.0\%), or every three months (48 participants, 19.4\%). Smaller groups reported different opinions: 23 participants (9.3\%) preferred changing passwords once a month, 2 participants (0.8\%) once a week, and 5 participants (2.0\%) every day. In contrast, 34 participants (13.7\%) felt that the frequency of password changes does not matter, while 16 participants (6.5\%) believed passwords never need to be changed. \\

A study by Habib et al. found that when password expiration policies are enforced, most users tend to modify their existing passwords rather than create entirely new ones, with fewer than a quarter choosing to generate completely new passwords\cite{219404}. A study by Zhang et al. revealed that when users update their passwords in predictable ways, the passwords become highly vulnerable, as attackers can exploit these patterns to crack them more easily\cite{zhang2010security}. In our study, 189 participants (76.2\%) only make small changes to their existing passwords rather than creating entirely new ones. Only 24 participants (9.7\% ) create entirely new passwords and 34 participants (13.7\%) base their  approach password change on the importance of the account. 1 person misunderstood the question and opted for the Other option. \\

Research has shown that a large proportion of users reuse the exact same passwords across multiple accounts and platforms\cite{hanamsagar2018leveraging,ur2015added}.
In our study, 231 participants (93.1\%) reported that they reused passwords, 8 participants (3.2\%) said that they did not, and 9 participants (3.6\%) indicated that they sometimes reused passwords depending on the importance of the website. Participants who reuse passwords frequently cited memorability, convenience, and a sense of perceived security as the primary reasons for doing so. Many also reported feeling burdened by the large number of passwords they must manage, which often leads them to favor convenience over stronger security practices. Participants who avoid reusing passwords tend to place a strong emphasis on security and privacy. Their primary concern is minimizing the risk that attackers could gain access to multiple accounts if one password were to be compromised. Participants who only occasionally reuse passwords tend to base their decisions on the perceived importance or sensitivity of an account. For example, they are more likely to create unique and stronger passwords for high-risk accounts, such as online banking, while reusing passwords for accounts they consider less critical, such as social media platforms.  A similar pattern was observed in the study by Hanamsagar et al., where they found that participants used passwords for important sites that were, on average, 1–2 characters longer and 20 times stronger than those for less important sites \cite{hanamsagar2018leveraging}. Wash et al. found that strong passwords tend to be reused more often \cite{wash2016understanding}. When asked whether passwords should be reused, a majority of 134 participants (54.2\%) responded "no" and 78 participants (31.4\%) answered "It depends". A minority of 36 participants (14.5\%) answered "yes". \\

Our participants' current password managing habits are ineffective, as reflected in their responses about password reset due to forgetfulness: 28 participants (11.3\%) never have to reset, 167 participants (67.3\%) occasionally reset, 43 participants (17.3\%) reset somewhat frequently, and 10 participants (4.0\%) reset very frequently. Similar to Gao et al.'s study where found that 59\% of their participants reset their passwords about once or several times per year\cite{gao2018forgetting}. Participants were also asked how often they forget which password belongs to a specific account. Among the respondents, 107 participants (43.0\%) reported experiencing this occasionally, while 72 participants (29.0\%) said it occurs somewhat frequently. Furthermore, 40 participants (16.0\%) said that they very frequently forget their passwords, while only 29 participants (12.0\%) said that they never experienced the issue.

\subsection{Insecure Practices}
A study by Ur et al. discovered that users' lack of understanding of the scale of potential attacks appears to be a fundamental cause of weak passwords \cite{ur2016users}. The same reasoning might explain why people often use poor security practices, such as using public Wi-Fi, not using multi-factor authentication, and failing to monitor leaked passwords. In our study, 188 participants (75.8\%) use public Wi-Fi, 38 participants (15.3\%) use public Wi-Fi depending on the situation and 22 participants (8.9\%) do not use public Wi-Fi. Regarding multi-factor authentication (MFA), 82 participants (33.3\%) reported that they enable it whenever the option is available. Meanwhile, 95 participants (38.3\%) stated that they use MFA only when it is mandatory, and 43 participants (17.3\%) use it specifically for important accounts. Additionally, 24 participants (9.6\%) said they were unfamiliar with MFA or did not know how to set it up, while 4 participants (1.6\%) reported that they chose not to use MFA at all. Our findings are consistent with Ion et al.'s study, which revealed that many participants hesitated to use multi-factor authentication. The study concluded that increased awareness of multi-factor authentication is necessary \cite{ion2015no}. Furthermore, only a tiny fraction (31 participants, 12.5\%) use auto-generated passwords, and only 27 participants (10.9\%) use services to track leaked passwords.\\

A study by Pearman et al. found that passwords used for government websites are reused across fewer domains, whereas passwords used for shopping and job search sites are reused more frequently\cite{pearman2017let}. Our study showed similar behavior among our participants. When asked which accounts they protect with the strongest passwords, participants most commonly mentioned bank, school, work, email, and government accounts. Some participants also reported using strong passwords for Wi-Fi networks, insurance accounts, military accounts, personal devices, airline accounts, Venmo accounts, and password manager accounts. However, two participants admitted that they use the same password across all their important accounts, even though they are aware that this is not a secure practice. \\

\subsection{Managing Passwords}
In today's digital landscape, individuals face a challenging dilemma in balancing the convenience of digital solutions with adherence to robust security practices. People have their own unique ways of managing their passwords. A study was conducted by Pearman et al. to evaluate the effective use of password managers concluded that individuals often do not utilize these tools optimally \cite{pearman2019people}. Most individuals primarily relied on memorization or password resets for managing their credentials \cite{219404,choong2014united}. Others utilized password management features embedded within their browsers or operating systems, such as those provided by Apple, Google, or Microsoft. Only a small portion of participants opted for standalone password management tools \cite{pearman2019people}. Our study yielded similar results where 194 participants (78.2\%) answered that they do not use a password manager and 54 participants (21.7\%) answered that they do use a password manager. Among the people who use password managers, 5 participants (8.06\%) use 1Password, one participant (1.61\%) said that they use "PW app available for Iphones" , 8 participants (12.9\%) use "Google's password manager", 16 participants (25.8\%) use iCloud password manager, 1 participant (1.61\%) uses KeePass, 2 participants (3.22\%) use Microsoft password manager, 16 participants (25.8\%) use Google Chrome, 2 participants (3.22\%) use LastPass, 2 people (3.22\%) use Bitwarden, 1 participant (1.61\%) uses Atlas Genius, 1 participant (1.61\%) uses iCloud Keychain, 1 participant (1.61\%) uses Opera. Our study confirms that individuals who use password managers tend to exhibit stronger security practices than those who do not. These users are more likely to monitor for leaked passwords actively, utilize auto-generators to create strong and unique passwords for each account, and only a notably smaller percentage rely on writing passwords down on paper. \\

Other studies have shown that many users keep records of their passwords either on paper or in an electronic form \cite{choong2014united,stobert2018password,shay2010encountering, komanduri2011passwords}. In our study, 149 people (60.1\%) answered that they do not write passwords down on paper, and 99 people (39.9\%) answered that they do write down passwords on paper. We tried to determine if a person's field of study affects their password habits, specifically their habit of writing down passwords on paper. We found that individuals in Education and Human Development are the most likely to write down passwords. Interestingly, our data revealed that having a technical background does not necessarily lead to better password security practices, as over 40\% of engineers in our study also admitted to writing their passwords on paper. \\

Pearman et al. found that individuals not using password management tools often engage in risky password practices, such as frequently reusing passwords and avoiding using unique ones\cite{pearman2019people}. A similar behavior was observed among participants who relied on built-in browser password tools, even though their password managers provided access to password generators. In contrast, participants who used standalone password managers reported regularly generating random and unique passwords for their accounts and consistently avoiding password reuse\cite{pearman2019people}. Many users of standalone password managers considered the password generator feature convenient, while most users of built-in password managers were unaware that such a feature existed\cite{pearman2019people}.\\

In our study, 230 people (92.7\%) said they do not use password generators, and 18 people (7.2\%) said they do use password generators. Among the participants who did say yes, 6 people out of 18 (33.3\%) said that they use Apple, 6 people out of 18 (33.3\%) said that they use Google, 1 person out of 18 (5.6\%) said that they use KeePass, 2 people out of 18 (11.1\%) said that they use Last Pass, 1 person out of 18 (5.6\%) said that they use 1Password, 1 person out of 18 (5.6\%) said that they use Microsoft and 1 person out of 18 (5.6\%) said "the one provided by the website". In contrast to Pearman et al.'s findings, our study reveals that individuals who use password generators predominantly rely on those provided by built-in password managers. \\

In our study, the majority of participants 221 people (89.1\%) do not actively track leaked passwords, which could pose a significant risk to their online security. Among those who track leaked passwords (27 participants, 10.9\%), there is a clear preference for using services provided by major platforms like Google and Apple, with a smaller number relying on Microsoft's services, LifeLock, LastPass, or family members. The overall low usage of tools for tracking leaked passwords, combined with the previously noted low adoption of password managers, suggests a broader need for increased awareness and education on comprehensive digital security practices. Our study observed an interesting trend among individuals who use password auto-generators and password managers. Regardless of whether they used auto-generators, participants commonly reported that they "occasionally" forgot which passwords corresponded to specific accounts, leading to password resets due to forgetfulness. The second most frequent response was “somewhat frequently,” suggesting that users may either lack familiarity with effectively using auto-generators connected to password managers or may be using them in unintended ways. Similar observations were made by Pearman et al., where they concluded that individuals often do not utilize these tools optimally \cite{pearman2019people}. 

\section{Discussion}
Our findings show that although many participants recognize the importance of using strong passwords, updating them regularly, and avoiding password reuse, they often prioritize convenience and ease of remembering passwords over security. The disconnect aligns with behavioral economic theories such as present bias, hyperbolic discounting, and status quo bias, which help explain why users prioritize short-term ease over long-term security. In this section, we analyze these behavioral patterns, explore the factors driving security trade-offs, and discuss potential strategies to bridge the gap between security awareness and practice. \\

\subsection{Security-Usability Dilemma}

The tension between security and convenience remains a central theme in password management. Our findings, along with prior research, demonstrate that while users generally acknowledge the risks of insecure passwords, they frequently make trade-offs that prioritize usability over robust security. A significant portion of participants knowingly engage in insecure practices to facilitate easy recall. Many participants admitted to using personal information in their passwords, which show that users often choose meaningful words or names, with minor modifications such as numbers or special characters \cite{shay2010encountering, hanamsagar2018leveraging}. Password complexity also appears to be selectively applied based on perceived account importance. Participants reported using longer, more complex passwords for critical accounts while opting for simpler passwords for social media or entertainment platforms. The “tiered” approach, also noted by Pearman et al. and Wash et al., indicates a pragmatic response to password overload \cite{pearman2019people, wash2016understanding}. Furthermore, password reuse remains alarmingly common. In our survey, 93.1\% of participants admitted to reusing passwords across different accounts, aligning with previous studies by Hanamsagar et al. and Wash et al., which found a similar propensity for reuse, with only minor modifications \cite{hanamsagar2018leveraging, wash2016understanding}. The reluctance to change passwords unless required highlights another dimension of convenience trumping security. Over half of our participants (63.3\%) update their passwords only when mandated by a system, mirroring similar findings by NIST and Habib et al. \cite{choong2014united, 219404}. \\

Difficulty managing multiple credentials further influences risky behavior. Many participants expressed feeling overwhelmed by the number of passwords they need to remember, leading them to adopt unsafe strategies. Although password managers are a recommended solution, only 21.7\% of participants use them. Similar results have been seen in Pearman et al.’s findings that password managers remain underutilized due to concerns about trust, lack of awareness, and perceived complexity \cite{pearman2019people}. Our results echo Hanamsagar et al. and Creese et al., who found that users’ risk perceptions do not necessarily lead to stronger passwords \cite{hanamsagar2018leveraging, creese2013relationships}. Ultimately, our findings reveal mixed adherence to secure password practices. Although many participants possess a solid understanding of safe practices, practical challenges, such as memorability and time constraints, frequently take precedence. The gap between knowledge and daily practice underscores the urgency of solutions that bridge security and convenience. Encouraging password managers, multi-factor authentication, and auto-generated passwords can mitigate risks without imposing excessive burdens on users. As the digital landscape expands, striking a balance between strong security measures and user-friendly tools will foster more consistent adoption of best practices. 

\subsection{Understanding Users Password Behaviors Through the Privacy Paradox}
Our study highlights a manifestation of the privacy paradox, the well-documented discrepancy between individuals' stated privacy concerns and their actual online security behaviors \cite{kokolakis2017privacy, barth2017privacy}. While participants demonstrated awareness of password security risks but their actual practices often contradicted the knowledge. For example, 72.1\% of participants acknowledged that using personal information in passwords is insecure, yet 57.2\% admitted to doing so. Similarly, while 53.6\% of participants believed passwords should not be reused, a striking 93.1\% still reused passwords. These figures suggest that while participants understand best practices, convenience shapes their behaviors more than security concerns. Research in the privacy paradox phenomenon, indicates that users struggle to balance security with usability, often opting for practices prioritizing memorability over robust security measures \cite{barth2017privacy}. The paradox extends to multi-factor authentication (MFA) and password management tools. Despite MFA being a widely recommended security measure, only 33.0\% of participants enabled it whenever available, and 9.6\% were unfamiliar with it. Likewise, 78.2\% of participants did not use a password manager, with many still relying on insecure methods such as writing passwords down on paper. The reluctance to adopt these security-enhancing tools suggests a gap between awareness and action, potentially due to perceived inconvenience or lack of trust in password management solutions \cite{kokolakis2017privacy}. Prior studies on the privacy paradox suggest that the discrepancy stems from habitual behaviors, risk habituation, and cognitive biases that influence decision-making in online security contexts \cite{barth2017privacy}. Strategies such as habit formation, nudging mechanisms, and improved usability of security tools could help users align their password practices with their stated security concerns.\\

\subsection{Factors Affecting Password Security Decisions}
Understanding the factors that influence people's decisions about passwords is essential to promote better security practices. Behavioral economic concepts such as hyperbolic discounting and present bias provide valuable insights into these decisions, particularly in the context of balancing immediate convenience with long-term security.\\

Hyperbolic discounting suggests that users disproportionately value immediate convenience over the future benefit of stronger security\cite{laibson1997golden}.
Through hyperbolic discounting we can see how impulsive behavior and prioritizing immediate rewards over long-term benefits can lead to poor decision making \cite{samson2016behavioral}. Our study found that many participants preferred passwords between eight and twelve characters long, even though they understood that longer passwords offer stronger security. Password reuse was also common, as managing unique passwords for multiple accounts was considered difficult and inconvenient. They still continued to use personal information and predictable password modifications despite being aware of the associated risks. \\

Status quo bias can explain that users often prefer familiar or existing security habits over adopting more secure alternatives, even when better options are available. The resistance to change may explain why many users are reluctant to use password managers or enable multi-factor authentication \cite{samuelson1988status, baumeister2001bad}. While participants recognize the added security benefits, the immediate “cost” of additional steps and potential inconvenience often outweighs the more distant threat of account compromise. Similarly, the small fraction who use auto-generated passwords or track leaked passwords underscores the tendency to discount prospective security risks, prioritizing short-term ease of use. \\

Present bias further explains why even those who know that shorter simpler passwords, personal details in passwords and common words in passwords increase vulnerability still prioritize convenience and familiarity, choosing to “pay” the immediate lower cost of a faster recall, which means less mental effort rather than protecting against a more distant and uncertain security threat. Some people understand that avoiding dictionary words and personal details is safer, yet they continue to use them, assuming that their specific or obscure details offset the risk. Meanwhile, others adopt a conditional approach which are stricter measures for higher-security accounts, suggesting they weigh the trade-offs in real time, often favoring immediate ease unless the perceived stakes are sufficiently high. Overall, the discrepancies between users’ theoretical knowledge of best practices and their reported behaviors highlight how the pull of present convenience can overshadow long-term security considerations. \\

\subsection{Limitations}
The study has several limitations that should be acknowledged. First, our data was collected from a single city and university, which may limit the generalization of the findings to other regions. Additionally, there was a notable gender disparity among participants, potentially introducing bias into the results. The sample also consisted predominantly of individuals aged 18 to 25, restricting insights into password security habits across a broader age range. The study relies on self-reported survey data, which introduces potential biases such as social desirability bias, recall bias, and question misinterpretation. Using behavioral economic theories may not accurately represent the complex nature of human behavior. Theories such as hyperbolic discounting, present bias and status quo bias may oversimplify human behavior. These theories do not fully account for individual cognition, personality, and context differences. People may be susceptible to biases based on their unique experiences, cultural background, or cognitive abilities. In real-world scenarios, numerous contextual factors such as social influence, workplace policies, or technical constraints can impact password-related behaviors, potentially confounding the theoretical predictions. While these theories can help explain people's security decisions, they may not directly translate to effective interventions. Designing practical solutions to improve password practices may require a more nuanced understanding of user motivations and environmental influences beyond what these theories provide. The research applies theoretical frameworks without incorporating quantitative modeling to validate these concepts. Without formal computation or statistical validation, the strength of the theoretical claims remains unverified. \\

\section{Conclusion}
The study reinforces previous research showing that although users understand recommended password security practices, they often prioritize convenience and ease of use over strong security behaviors. Participants demonstrated awareness of secure password habits but continued to engage in risky practices such as reusing passwords, relying on personal information, and making only minor modifications when updating passwords. These behaviors reflect psychological tendencies such as present bias and hyperbolic discounting, where immediate usability and memorability are valued more highly than long-term security benefits. The research further showed that users apply different levels of password security depending on how important they perceive an account to be. Stronger and more complex passwords were generally reserved for critical accounts such as banking or email services, while weaker passwords were commonly used for social media and entertainment platforms. The selective approach suggests that users often underestimate the risks associated with less important accounts, creating additional vulnerabilities across multiple online services. Another key finding was the reluctance to adopt password managers and multi-factor authentication despite their well-established security benefits. Concerns about usability, trust, and the effort required to learn new tools discouraged many participants from using these protective measures. As a result, many users continued relying on familiar but less secure password management habits. The study concludes that improving password security requires more than simply increasing user awareness. Effective solutions should focus on reducing cognitive effort and aligning security practices with natural human behavior. Encouraging the use of user-friendly authentication tools, simplifying security processes, and correcting common misconceptions about password management could help bridge the gap between security knowledge and real-world behavior. Ultimately, balancing usability with security remains essential for promoting safer online practices and strengthening digital security overall.

\bibliographystyle{plain}
\bibliography{usenix2025_soups_latex-template/usenix2025_SOUPS}


\section*{Appendix: Survey}
\vspace{5pt}
\subsection*{Demographics}
1. How old are you? (note: you must be over 18 to take this survey)
\begin{itemize}
  \item 18-25
  \item 26-39
  \item 40-55
  \item 55+
\end{itemize}
2. What is your gender?
\begin{itemize}
  \item Male
  \item Female
  \item Non-Binary
  \item Other
  \item Prefer not to say
\end{itemize}
3. What is your field of study (if you are a student) or area of work (if you are in the workforce)?
\begin{itemize}
  \item Computer Science \& Engineering
  \item Other Engineering (mechanical, electrical, civil, etc)
  \item Science/other STEM field
  \item Medicine
  \item Law
  \item Business
  \item Education
  \item Humanities
  \item Retail
  \item Food service/culinary
  \item Retired
  \item Prefer not to say

 \item Other
\end{itemize}
\subsection*{Personal Password Security}
1. Which of these looks about the same length as your average password?
\begin{itemize}
  \item \verb|s^9uN$f6| (8 characters)
  \item \verb |B&^s^9uN$f64| (12 characters)
  \item \verb |z#B&^s^9uN$f64NY4| (17 characters)
  \item \verb |M1z#B&^s^9uN$f64NY4leF| (22 characters)
  \item \verb |qpCXZM1z#B&^s^9uN$f64NY4leFRAPuA| (32 characters)
  \item It depends on what it's for
\end{itemize}
2. How often do you change your password?
\begin{itemize}
    \item Never
    \item Every day
    \item Once a week
    \item Once a month
    \item Every three months
    \item Every six months
    \item Every year
    \item Whenever I am required to
\end{itemize}
3. When you change a password, do you create an entirely different password or do you make a small change to the original password (ex: changing 'password123' to 'password124')? 
\begin{itemize}
    \item Entirely different password
    \item Small change to the original password
    \item Depends on what it's for
\end{itemize}
4. Do you ever log into your accounts on public Wi-Fi? 
\begin{itemize}
    \item Yes 
    \item No
    \item Only some accounts
\end{itemize}
5. Do you ever reuse passwords? 
\begin{itemize}
    \item Yes
    \item No
    \item Sometimes
\end{itemize}
6. Regarding your previous answer, why or why not? \\
7. Do you use multi factor authentication (MFA)?
\begin{itemize}
    \item Yes, whenever it's an option
    \item Sometimes, but only for very important accounts
    \item Only when it is required
    \item No, I'm not sure what it is or not sure how to set it up
    \item No, I choose not to
\end{itemize}
8. Do any of your passwords include a dictionary word? (Example of a password with a dictionary word: Card@2023)
\begin{itemize}
    \item Yes
    \item No
    \item Some
\end{itemize}
9. Regarding your previous answer, why or why not? \\
10. Do any of your passwords include personal information? 
\begin{itemize}
    \item Yes
    \item No
    \item Some
\end{itemize}
11. Regarding your previous answer, why or why not? \\
12. Does the security of your passwords depend on what they are used for? 
\begin{itemize}
    \item Yes
    \item No
\end{itemize}
13. Do you use auto-generated passwords? 
\begin{itemize}
    \item Yes
    \item No
\end{itemize}
14. Which accounts do you use the most secure passwords for?\\
\subsection*{Making Secure Passwords}
1. How long should a password be?
\begin{itemize}
    \item Shorter than 8 characters
    \item 8-12 characters
    \item 13-20 characters
    \item 21-32 characters
    \item 33+ characters
    \item It doesn't matter
\end{itemize}
2. How often should you change your password?
\begin{itemize}
    \item Never
    \item Every day
    \item Once a week
    \item Once a month
    \item Every three months
    \item Every six months
    \item Every year
    \item It doesn't matter
\end{itemize}
3. What are some ways to make a password complex?\\
4. Should your password include any dictionary words in it?
\begin{itemize}
    \item Yes
    \item No
    \item It depends
\end{itemize}
5. Regarding your previous answer, why or why not?\\
6. Is it safe to use personal information, such as your birthday or name, in a password?
\begin{itemize}
    \item Yes
    \item No
    \item It depends
\end{itemize}
7. Regarding your previous answer, why or why not?\\
8. Should you reuse passwords? 
\begin{itemize}
    \item Yes
    \item No
    \item It depends
\end{itemize}
9. Regarding your previous answer, why or why not?\\
\subsection*{Keeping Track of Passwords}
1. Do you use a password manager?
\begin{itemize}
    \item Yes
    \item No
\end{itemize}
2. If so, which one? \\
3. Do you use any services that track leaked passwords? 
\begin{itemize}
    \item Yes
    \item No
\end{itemize}
4. If so, which one? \\
5. Do you write your passwords down on paper anywhere? 
\begin{itemize}
    \item Yes
    \item No
\end{itemize}
6. Do you share your passwords with anyone else? If so, who? \\
7. Do you use a generator to come up with passwords? If so, which one? \\
8. How often do you have to reset a password due to forgetfulness? 
\begin{itemize}
    \item Very frequently
    \item Somewhat frequently
    \item Occasionally
    \item Never
\end{itemize}
9. How often do you forget which passwords belong to which accounts of yours? (For example, mixing up your bank and email passwords) 
\begin{itemize}
    \item Very frequently
    \item Somewhat frequently
    \item Occasionally
    \item Never
\end{itemize}

\end{document}